\documentclass{svproc}
\usepackage[hidelinks]{hyperref}
\usepackage{amsmath}
\usepackage{wrapfig}
\usepackage[square,sort,comma,numbers]{natbib}
\usepackage[utf8]{inputenc}
\usepackage{float}
\usepackage{array} 
\usepackage{siunitx}

\usepackage{url}
\usepackage{graphicx}

\usepackage{todonotes}

\begin{document} 
\presetkeys{todonotes}{noinline, color=teal}{}

\tikzset{/tikz/notestyleraw/.append style={text=white}}

\mainmatter              
\title{Source mass characterization in the ARIADNE axion experiment}
\titlerunning{Source-mass characterization in the ARIADNE axion experiment}  
%
%
\authorrunning{C. Lohmeyer for the ARIADNE collaboration} 
%
\tocauthor{C. Lohmeyer for the ARIADNE collaboration} 

\newcommand{\NU}{{ Center for Fundamental Physics,Department of Physics and Astronomy, Northwestern University}}
\newcommand{\PI}{Perimeter Institute, 31 Caroline St N, Waterloo, ON N2L 2Y5, Canada}
\newcommand{\IU}{Department of Physics, Indiana University, 727 E Third St, Bloomington, IN 47405}
\newcommand{\SUA}{Department of Physics, Stanford University, 382 Via Pueblo, Stanford, CA 94305}
\newcommand{\SUB}{Department of Physics and Applied Physics, Stanford University, 382 Via Pueblo, Stanford, CA 94305}

\newcommand{\LANL}{Los Alamos National Laboratory, Los Alamos, NM 87545}

\newcommand{\CAPA}{Center for Axion and Precision Physics Research, IBS, 193 Munji-ro,  Daejeon, South Korea, 34051, Department of Physics, KAIST, 291 Daehak-ro, Daejeon, South Korea, 34141}

\newcommand{\CAPB}{Center for Axion and Precision Physics Research, IBS, 193, Munji-Ro,  Daejeon, South Korea, 34051}

\newcommand{\KRISS}{KRISS, 267 Gajeong-ro, Yuseong-gu, Daejeon 34113, Republic of Korea}

\newcommand{\PTB}{Physikalisch-Technische Bundesanstalt, Abbestraße 2, 10587 Berlin, Germany}

\author{C. Lohmeyer for the ARIADNE Collaboration\\
N. Aggarwal\inst{1},  A. Arvanitaki\inst{2}, A. Brown\inst{3} 
A. Fang\inst{4},
A.A. Geraci\inst{1},
A. Kapitulnik\inst{4}, D. Kim\inst{5}, Y. Kim\inst{5},
I. Lee\inst{3}, Y.H. Lee\inst{6}, E. Levenson-Falk\inst{7}, C.Y. Liu\inst{3}, C. Lohmeyer\inst{1}, J.C. Long \inst{3},
S. Mumford\inst{7}, 
 A. Reid\inst{3}, 
Y. Semertzidis\inst{8}, Y. Shin\inst{8}, J. Shortino\inst{3}, E. Smith\inst{9}, W.M. Snow\inst{3},   
E. Weisman\inst{3}
\linebreak
\linebreak
A. Schnabel\inst{10}
L. Trahms\inst{10},
J. Voigt\inst{10}}
\institute{\NU \and \PI \and \IU \and \SUB \and \CAPA \and \KRISS \and \SUA  \and \CAPB \and \LANL \and \PTB}

\maketitle

\begin{abstract}
The Axion Resonant InterAction Detection Experiment (ARIADNE) is a collaborative effort to search for the QCD axion using nuclear magnetic resonance (NMR), where the axion acts as a mediator of spin-dependent forces between an unpolarized tungsten source mass and a sample of polarized helium-3 gas. Since the experiment involves precision measurement of a small magnetization, it relies on limiting ordinary magnetic noise with superconducting magnetic shielding. In addition to the shielding, proper characterization of the noise level from other sources is crucial. We investigate one such noise source in detail: the magnetic noise due to impurities and Johnson noise in the tungsten source mass.
\keywords{Axion, ARIADNE, Nuclear magnetic resonance, 3He, SQUID, Interaction, Fifth-force, Superconductor, Magnetic impurity}
\end{abstract}
\section{Introduction}
 Axions or axion-like particles arise in many theories beyond the standard model \cite{minaandy,axion, ALPS}. One such particle, the Peccei-Quinn (PQ) axion, was postulated in the 1970s as a solution to the strong CP problem in quantum chromodynamics (QCD) \cite{axion}. The strong CP problem arises from experimental limits on the electric dipole moment (EDM) of the neutron being magnitudes smaller than would be expected theoretically. This discrepancy is usually parameterized by an angle $\theta_{\rm QCD}$, which is generally expected to be of order unity under QCD, but is observed to be less than $10^{-10}$ by searches for the neutron EDM \cite{mercuryEDM,neutronEDM}.  In addition to its usefulness in explaining the smallness of the $\theta_{QCD}$ angle \cite{axion,moody1984ba}, the axion is also considered to be a promising dark matter candidate  \cite{axionCDM,axionCDM2}. The parameter space for the QCD axion in particular can be parameterized by mass, which is inversely proportional to the PQ symmetry breaking scale $f_a$ \cite{axion}. The mass range has an upper bound of $m_A < 10 $ meV which is implied by astrophysical limits \cite{PDG} and a lower bound dependent upon cosmological theories of inflation where $m_A > 1$ $\mu$eV \cite{axionCDM,axionCDM2} for models with the energy scale of inflation above the PQ phase transition. The mass of the QCD axion can be significantly lighter in models with low energy-scale inflation \cite{axionCDM,axionCDM2}. ARIADNE aims to search for an axion field in the mass range $1$ $\mu$eV $< m_A <6$ meV that couples to nucleons via their spin \cite{minaandy}. Other axion experiments and proposals for new experiments probe for axions in different mass ranges, using a variety of couplings (i.e. to photons, electrons, nucleons), and unlike our approach, search for interactions directly with the hypothetical cosmic dark matter axion field.  \cite{ADMX, casper, orpheus, MADMAX, DMRadio, HAYSTAC, ABRACADABRA, QUAX, LCcircuit}. A comparison can be found in reference \cite{proceed2}, showing ARIADNE's potential to fill in an uncharted and complementary region of the QCD axion parameter space.

ARIADNE searches for the QCD axion from the perspective that it mediates short-range spin-dependent forces between two nucleons. The interaction energy can be written as \cite{minaandy} 
\begin{equation}
    U_{sp}(r)=\frac{\hbar^2 g_s^N g_p^N}{8 \pi m_f}\left( \frac{1}{r \lambda_a}+\frac{1}{r^2}\right) e^{-\frac{r}{\lambda_a}} \left(\hat \sigma \cdot \hat r \right), \label{eq:1}
\end{equation}
where $g_s$ and $g_p$ are the monopole and dipole coupling constants, respectively, $\lambda_a$ is the Compton wavelength of the axion, $m_f$ is the fermion mass, $\hat{\sigma}$ is the fermion spin, and $\vec{r}$ is the vector between the source mass and spin. 
This coupling to the fermion spin allows us to define an effective or fictitious magnetic field. In order to detect this field, we plan to use an NMR technique. An ensemble of nuclear spins in gaseous $^{3}$He will be hyperpolarized by metastability-exchange optical pumping (MEOP). An external magnetic field B$_{ext}$ 
will be applied to the nuclear spins to set their Larmor precession frequency at approximately $100$ Hz. A dense, nominally non-magnetic source mass is modulated in close proximity to the $^{3}$He sample resulting in a modulation of the effective transverse magnetic field. When the frequency of the field modulation matches the Larmor frequency, the fermion spins will be resonantly driven into precession \cite{minaandy}.  This results in a transverse magnetization which can be detected using a SQUID magnetometer. A full description of the experiment can be found in Ref. \cite{proceed2}. 

\begin{figure}
    \centering
    \includegraphics[width=\textwidth]{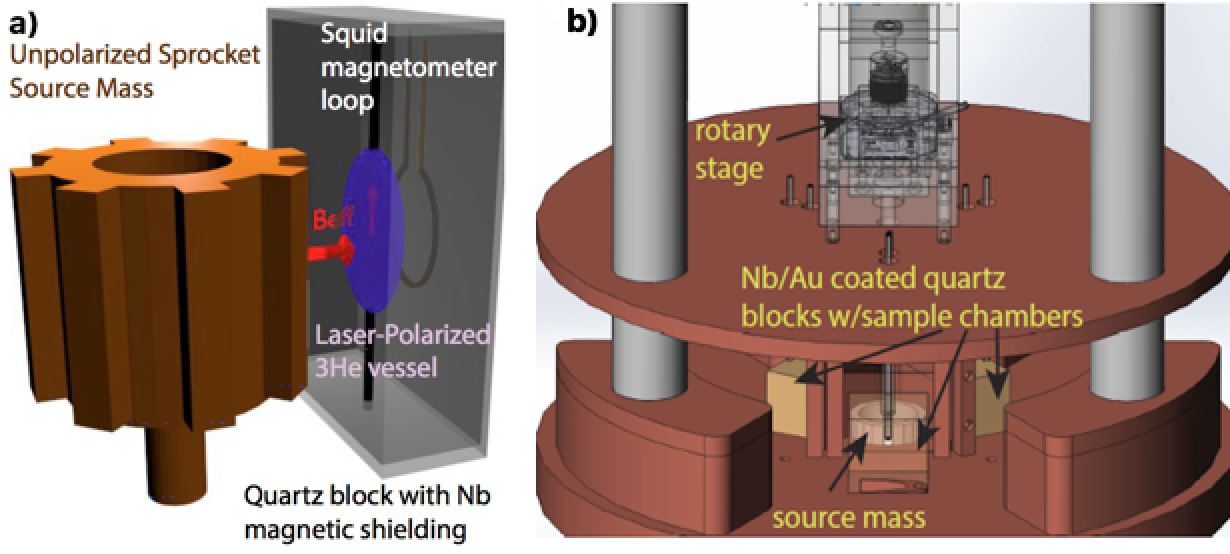}
    \caption{\textbf{a)} Illustration of the experiment concept: the unpolarized tungsten source mass is placed near a quartz block with a thin film of niobium shielding. Inside the block is  the spheroidal vessel housing $^3$He and the SQUID magnetometer.  \textbf{b)} Detailed schematic of the bottom region of the cryostat with three ${^3}$He sample vessels surrounding the tungsten sprocket. The sample cells will be connected to a copper stage at 4K.}
    \label{fig:sourcesample}
\end{figure}

The source mass will be a tungsten sprocket with 11-fold symmetry, as shown in Figure \ref{fig:sourcesample}a. The source mass will be rotated such that the ``teeth'' transit the $^{3}$He cell at the Larmor frequency 
The receptacle which houses the $^3$He sample will be a spheroidal shaped cavity in a piece of fused quartz. Three $^3$He sample chambers will be positioned around the sprocket so that cross-correlation among them can be used to suppress backgrounds (see Figure \ref{fig:sourcesample}b). The average gap between the tungsten source mass and the $^{3}$He sample is kept below $200$ $\mu$m, (within the Compton wavelength of the axion, as determined by the targeted mass range) in order to detect the pseudo-magnetic field.

Since the axion signal manifests as a magnetization of the nuclear spins, we must shield the spin sample from electromagnetic backgrounds. In order to achieve this, a superconducting magnetic shielding will be deployed.
This shield is formed by sputtering a $\sim 1$ $\mu$m layer of niobium onto the surfaces of the quartz chambers. The shield is anticipated to suppress magnetic backgrounds for sufficiently low magnetization and contamination of the source mass. A thorough characterization of a source mass prototype is the main focus of this paper. Here we present a description of the source mass fabrication and specifications as well as residual magnetization measurements. Application of a large external magnetic field to the prototype resulted in a residual permanent magnetization, indicating the presence of magnetic impurities. However, this magnetization could be reduced by an order of magnitude by de-gaussing. The remaining magnetization as well as Johnson noise ultimately determines the magnetic shielding requirements for the thin film Nb shield. The results obtained thus far are sufficiently low for the anticipated shielding factor.

\section{Source Mass Fabrication and Characterization}

\subsection{Prototype source mass}
The raw tungsten for the source mass is quoted to be greater than $99.95\% $ pure \cite{midwesttungsten}. This material was chosen for its high nucleon density as well as its low magnetic susceptibility.  The sprocket was machined using wire electrical discharging machining \cite{wireEDM}. The sprocket contains 11 teeth equally spaced around its circumference. The teeth ensure that the source drive frequency differs from the frequency of modulation in order to isolate vibrational backgrounds at the rotation frequency. The outer radius is 19 mm at the teeth and 18.8 mm between teeth. 
Thus, the source mass modulation amplitude is 200 microns as determined by the target axion mass range. The sprocket radii were machined with a dimensional tolerance of $\pm 10$ microns in order for the modulation of the axion signal to be uniform. The sprocket thickness is 1 cm in order to fully subtend the ${^3}$He sample cell. 

\begin{figure}
 \centering
    \includegraphics[width=\textwidth]{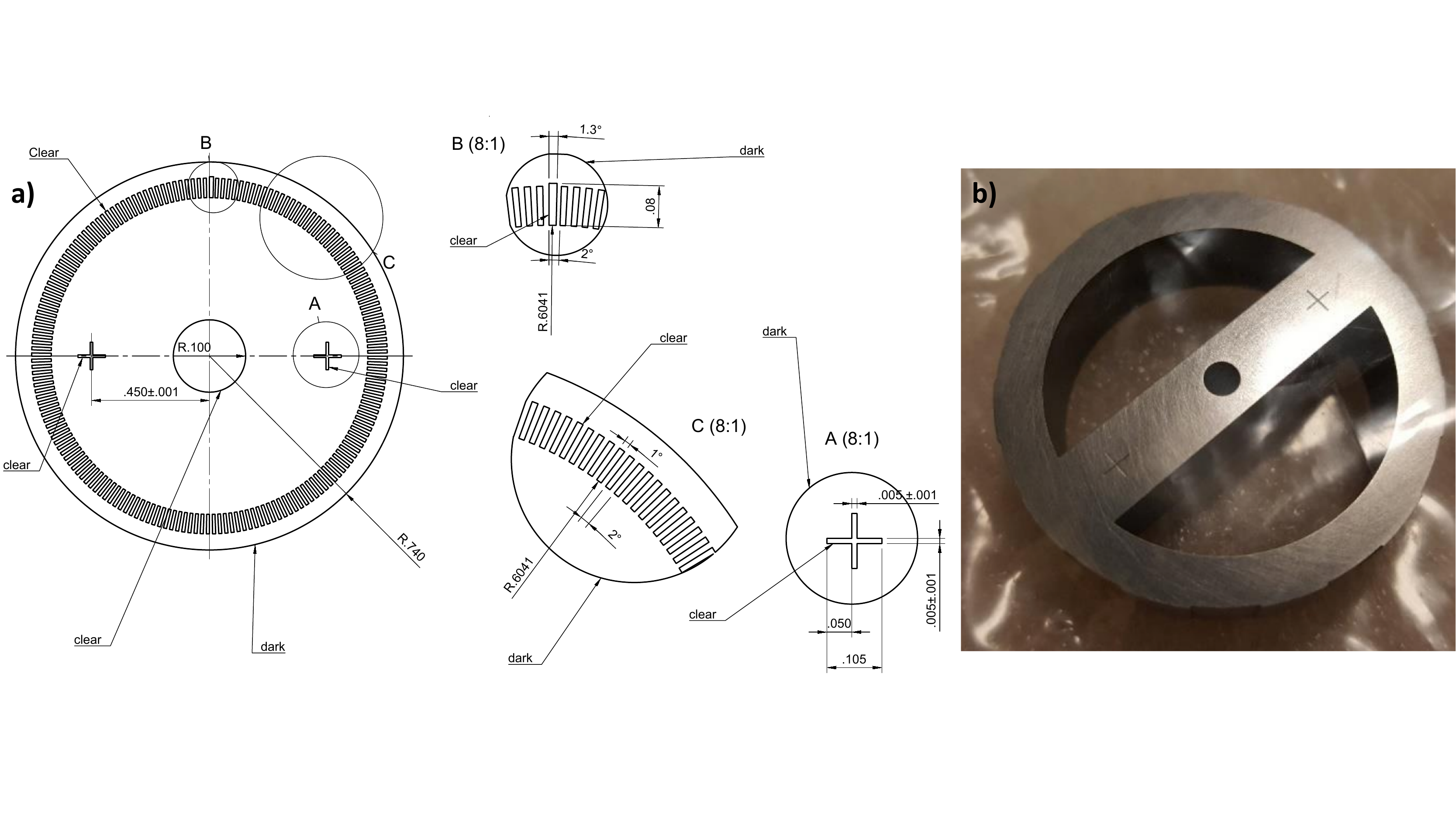}
   \begin{flushleft}
   \caption{\label{fig:SpinSpeed}} \textbf{a)} Schematic detailing the pattern that will be applied to the surface of the sprocket in order to monitor the rotational speed. \textbf{b)} A photograph of the prototype tungsten sprocket. 
   \end{flushleft}

\end{figure}
 The sprocket will need a high level of radial stability as it will be placed 50 $\mu$m from the surface of the quartz vessel that houses the helium sample. A fiber optic interferometer system will be employed to monitor a laser reflected off the bottom surface of the sprocket and assess both the speed stability and ``wobble" of the sprocket.  The surface finish of the sprocket will need to be highly reflective in this respect.  The top and bottom flat surfaces were finished with 220 and 440 grit abrasive sheet, giving an RMS surface roughness of $1.4$-$2.1$ $\mu$m as observed via profilometry measurements \cite{AlexBrownThesis}.
 In the experiment, the sprocket's rotational velocity needs to remain stable such that the modulation frequency stays matched to the Larmor precession. In order to monitor rotational velocity, a higher surface quality will be needed to facilitate the optical interferometry. The periodic velocity pattern will therefore be applied to an optically flat silicon disk epoxied to the bottom surface of the sprocket (Figure \ref{fig:SpinSpeed}a).

\subsection{Relevant magnetic backgrounds}

The sensitivity of the experiment to an effective magnetic field is expected to be limited by the transverse projection noise in the $^{3}$He sample, to the level of B$_{min}=$\SI{3e-19}{\tesla} (1000 s/T$_{2}$) $ T/\sqrt{Hz} $  where the T$_{2}$ is the $^{3}$He relaxation time. With a T$_{2}$ of 1000s and sufficient integration time \SI{e6}{s}, this sensitivity would be sufficient to probe the axion coupling (Eq.\ref{eq:1}) at the level below the current constraint set by the EDM limits. Thus it is important to keep other magnetic backgrounds below B$_{min}$

We anticipate the following to be the dominant sources of magnetic background: thermal Johnson noise, the Barnett effect, magnetic impurities in the source mass,  and the magnetic susceptibility of the source mass.

\emph{Johnson Noise} All electrical conductors have thermal motion of electrons which generate electronic noise, known as Johnson Noise \cite{johnsonnoise}. As a conductor, the tungsten sprocket will have an anticipated level of \SI{e-12}{\frac{T}{\sqrt{Hz}}} of Johnson noise at its expected temperature of $\sim$170 K. A magnetic shielding factor $\it{f}$ of $10^{8}$ lowers this background to \SI{e-20}{\frac{T}{\sqrt{Hz}}}.

\renewcommand{\arraystretch}{1.25}
\begin{table}[!hb]
\small
\begin{center}
\begin{tabular}{>{\raggedright}p{0.25\textwidth}p{0.35\textwidth}p{0.35\textwidth}}
  \hline
  \hline
Systematic Effect/ \linebreak Noise source & Background Level &    Notes \\
  \hline
Johnson noise& $10^{-20} (\frac{10^8}{f}) {\rm{T}}/\sqrt{{\rm{Hz}}} $ & $f$ is SC shield factor (100 Hz) \\
Barnett Effect& $10^{-22} (\frac{10^8}{f}) $ T & Can be used for calibration above 10 K \\
 Magnetic Impurities in Mass & $10^{-25}-10^{-17} (\frac{\eta}{1 {\rm{ppm}}}) (\frac{10^8}{f}) $ T & $\eta$ is impurity fraction \\ 
 Mass Magnetic  \linebreak Susceptibility & $10^{-22} (\frac{10^8}{f}) $ T & Assuming background field is $10^{-10}$ T \\
 &  & Background field can be larger if $f>10^8$ \\
  \hline
  \hline
  \label{tab:noise}
  \end{tabular}

\caption{\label{table1} Estimated source mass related systematic error and noise sources, as discussed in the text. The projected sensitivity of the device is $3 \times 10^{-19} (\frac{1000 {\rm{s}}}{T_2})^{1/2}$ T/$\sqrt{\rm{Hz}}$. Further discussion of systematic effects and noise levels can be found in Ref\cite{proceed2} }  
\end{center}
\end{table}

\emph{Barnett Effect} When an uncharged object is rotated on its axis, it will acquire a magnetization dependent on its magnetic susceptibility $\chi$ and frequency of rotation $\omega$ \cite{barnett}: 
\begin{equation} 
  M = \frac{\chi\omega}{\gamma} \label{eq:2}
\end{equation}
where $\gamma$ is the gyromagnetic ratio. For tungsten, $\chi =$ \SI{6.8e-5}{} \cite{magsusW} and the frequencies of rotation relevant for this experiment will be about 5 to 9 Hz. We expect a background magnetic field of $10^{-14}$T to be produced by the Barnett Effect. With a magnetic shielding factor of \SI{e8}{} the background noise is reduced to \SI{e-22}{T}. 

\emph{Magnetic Susceptibility} We anticipate a background level of \SI{e-22}{T} due to the magnetic susceptibility of the mass, assuming a background field of \SI{e-10}{T} and a shielding factor of \SI{e8}{}. 

\emph{Magnetic Impurities}
Iron contaminants can be present in ``pure" samples of tungsten metal. Assuming Fe impurities at a concentration of 1 ppm, we expect the residual magnetic field to be between \SI{e-17}{T} and \SI{e-9}{T}, depending on whether the moments are oriented randomly or perfectly aligned. Again, assuming a shielding factor of \SI{e8}, this background is reduced to \SI{e-25}{T} -- \SI{e-17}{T} which is above the specification but only for the worst case scenarios.

\subsection{Characterization Measurement Procedure}

\begin{figure}
  
   \includegraphics[width=\textwidth]{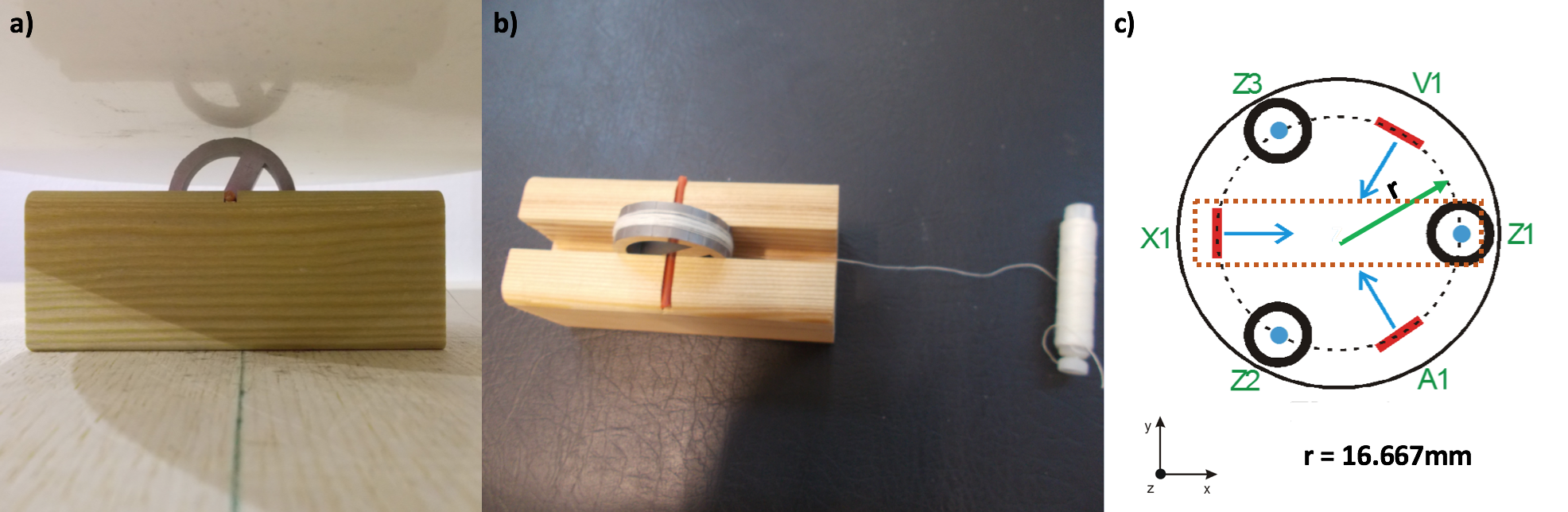}
   \caption{\label{fig:PTBsquid}} 
  \textbf{a)} Photograph of the tungsten sprocket in a wooden holder used for rotation positioned below the bottom surface of a dewar containing the SQUID measurement system. \textbf{b)} Photograph showing the wooden rotation assembly and string for manual rotation. \textbf{c)} A top view schematic of one of the 19 modules of the multi-channel SQUID system. SQUIDs Z1, Z2, Z3 are in the positive Z direction (out of the plane of the page, and thus vertical in photo (a)), and the SQUIDS X, V, and A are oriented to measure in the X and Y directions. The radius of the dotted black line is 16.667mm. The brown dotted rectangle is indicating the location and orientation of the rotor.
\end{figure}

In this section, we present results on the magnetization and magnetic impurity tests of the tungsten source mass and describe a future method for obtaining higher-precision results. Measurements were carried out at the Physikalisch-Technische Bundesanstalt (PTB) laboratory in Berlin. To enhance the sensitivity of these measurements to potential contaminants, the source mass was magnetized with a 30 mT neodymium magnet. The resulting magnetization was then measured in a magnetically-shielded room with a multi-channel superconducting quantum interference device (SQUID) system with the sprocket at room temperature. This SQUID system contains 19 modules where the the relevant plane to this measurement is the plane shown in \ref{fig:PTBsquid}c.
The blue dots and arrows indicate the orientation of each magnetometer pickup loop. Three of them are oriented to measure in the Z directions (Z1, Z2, Z3) and three oriented to measure in the X and Y directions (X1, V1, A1). These 6 signals can be used to assess of the direction and localization of magnetic contaminants. The brown dotted rectangle is indicting the location and orientation of the wheel. 

Static and rotational tests were performed. Using the measured field values from the SQUID sensor array, the net magnetization of the sample and inferred surface field were determined. A hard disk eraser generating a 500 mT magnetic field amplitude at 50 Hz was then used to demagnetize the sprocket via a usual decreasing AC-amplitude degaussing process. The magnetization was measured before and after degaussing.

The rotational tests were performed by placing the source mass in a wooden structure (Figure \ref{fig:PTBsquid}a) and placed below the dewar containing the SQUID system (Figure \ref{fig:PTBsquid}c) where the brown dotted rectangle depicts the orientation of the rotor. The rotation was done manually by string (Figure \ref{fig:PTBsquid}b) where the frequency of rotation was inconsistent. This rotational test was done before and after degaussing for overall magnetic field signal comparison. Further tests can be performed to analyze more closely the magnetic impurity orientations and locations. These tests will include measurements with QuSPIN \cite{quspin} optical magnetometers allowing for more control of the distance between sensor and sprocket, and a servo motor setup to provide rotation speed control.  This allows a more direct measurement of the near-surface fields, which are most relevant for the ARIADNE experiment.

\section{Results}
\begin{figure}[!ht]
        \centering
        \includegraphics[width=\textwidth]{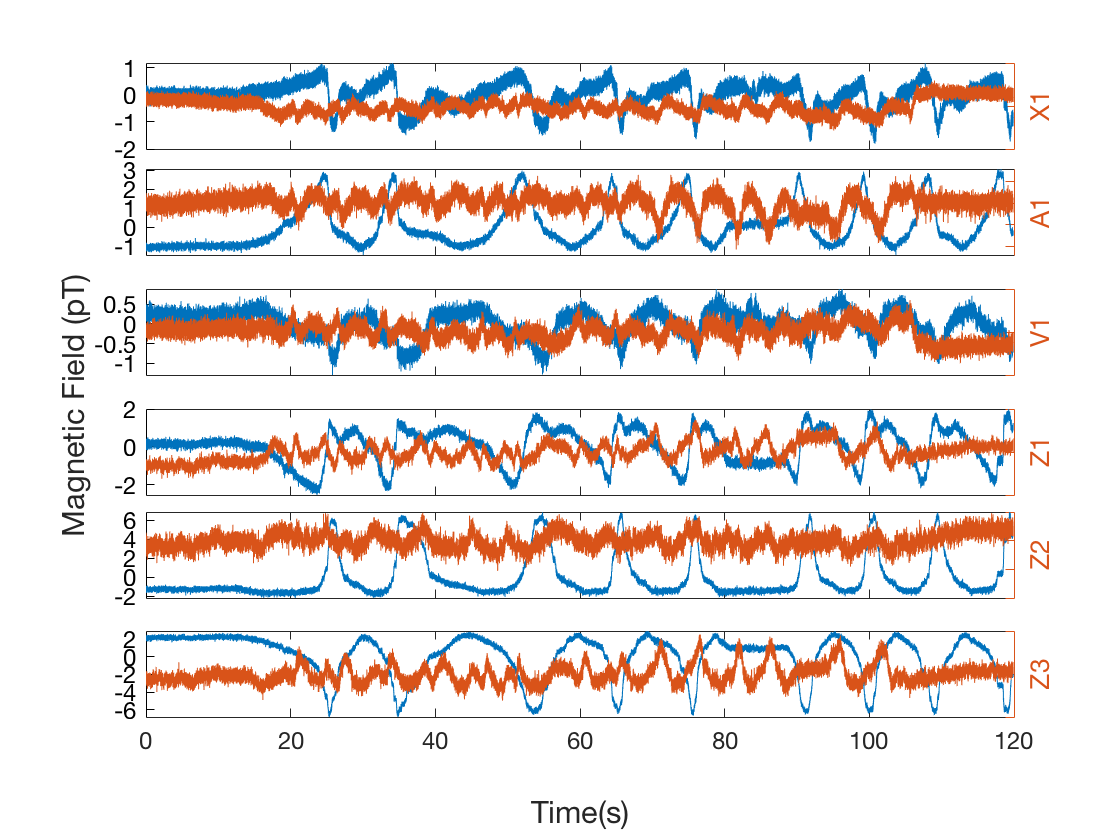}
        \begin{flushleft}
        \caption{\label{fig:PTB1}} Output from the 6 SQUID sensors (see \ref{fig:PTBsquid}c for the orientation) for the rotational test done before degaussing (in blue) and output after degaussing (in orange).
        \end{flushleft}
\end{figure}
The static test found that after magnetization the tungsten source mass exhibited a net magnetic dipole moment of about $20$ nAm$^2$ corresponding to an approximate surface magnetic field of about 300 pT, thus indicating that the source mass or its impurities can be slightly magnetized. 
The value of this dipole moment was deduced from the data collected from the 19-module squid array. 
After degaussing, the residual magnetic moment in the sample was reduced to a level which precluded a good fit to a magnetic dipole. The corresponding field signal at the sensors was reduced to approximately 2 pT at about a 3 cm distance. 

For the rotational tests, a string was pulled to rotate the wheel by approximately 14 full rotations over a time period of 120 seconds. Figure \ref{fig:PTB1} shows the output from the 6 SQUID sensors in the lowest level module.  The largest signals before degaussing have an amplitude around 8 pT$_{pp}$ (Figure \ref{fig:PTB1}). Here the separation between the sensors and the edge of the wheel is 30 mm $\pm$ 5 mm. After degaussing, the signals were reduced to approximately 2 pT$_{pp}$ (Figure \ref{fig:PTB1}). Additionally, the wheel was found to generate Johnson noise on the level of 1-1.5 pT$_{pp}$ for a bandwidth of about DC-100 Hz (250 Hz sampling rate).  Note that in Figure \ref{fig:PTB1}the 14 peaks present in the data (e.g. from sensor Z3 or Z2) correspond to the 14 manual rotations of the wheel. As the string was pulled manually, the time between revolutions of the wheel is not uniform. Future tests will employ a rotation mechanism with constant speed which will allow us to analyze the Fourier harmonics of the signal from the rotation and separate out any underlying signal which occurs at the 11-fold periodic pattern of the teeth in the wheel.   
Considering the data, we can roughly estimate that the residual field near the surface of the wheel after degaussing will be at or below the level of $~\sim 30$ pT. The data are consistent with a single, or possibly double, localized regions of contaminant, as opposed to uniformly distributed contaminants in the sprocket, since the modulated signal does not appear to be varying 11-fold upon each sprocket rotation. The residual signal at this level with 11-fold response per rotation can be bounded approximately at the few pT level, which would require a shielding factor in ARIADNE of approximately $10^8$ to achieve full design sensitivity.

We find these preliminary results to be encouraging, however further tests using the SQUID system and QuSPIN optical magnetometer system will be performed to obtain more quantitative results on the localization and orientation of any magnetic contaminants. Also a clearer measurement of the residual field from Johnson noise should be obtainable. These results will be compared with the background specifications for ARIADNE.

\section{Discussion and Outlook}

To summarize, we have tested a procedure to fabricate the source mass for the ARIADNE experiment and we have characterized its residual magnetism. Preliminary tests have shown that the impurities can be magnetized for detailed characterization, as well as demagnetized for eventual use in the experiment. 

Currently, we are developing methods to probe the field at a shorter distance from the source mass. This will also permit better control of its rotation speed as well as its position relative to the probe. This is an important step to ascertain if the current metal purity level and the fabrication procedure are acceptable for use in the ARIADNE experiment, or if there are any unexpected background sources. 

Besides the source mass, work is underway to construct the main cryostat that contains the liquid $^{4}$He cooling system,  the quartz holder for the sample $^{3}$He, and  superconducting niobium shield. Niobium coatings are being characterized for the critical temperature, shielding factor, and adhesion to a quartz surface.

\section{Acknowledgements}
We acknowledge the work of Alex Brown, who as an undergraduate at Indiana University produced the mechanical drawings for the tungsten sprocket described in this work, and conducted measurements to confirm that the raw tungsten material met the necessary magnetic specifications before the sprocket was cut.  Alex passed away in the fall of 2018 while beginning the graduate program in Physics at Florida State University.

 We acknowledge support from the U.S. National Science Foundation, grant numbers NSF PHY-1509176, NSF PHY-1510484, NSF PHY-1506508, NSF PHY-1806671, NSF PHY-1806395, NSF PHY-1806757.

%

\bibliographystyle{spmpsci_unsrt}
\bibliography{biblio}

\end{document}